\documentclass[11pt,preprint2]{aastex}

\usepackage{natbib}
\usepackage{epsfig}
\usepackage{prettyref}
\newcommand{\pref}{\prettyref}

 \newcommand{\be}{\begin{equation}}
 \newcommand{\ee}{\end{equation}}
 \newcommand{\ie}{\emph{i.e.}}
 \newcommand{\eg}{\emph{e.g.}}
 \newcommand{\kms}{\mbox{km\ \ensuremath{\rm{s}^{-1}}}}

\shorttitle{Density-enhanced gas and dust shells in IRC+10216}

\begin{document}

\title{Density-enhanced gas and dust shells in a new chemical model for IRC+10216}
\author{M. A. Cordiner \and T. J. Millar}
\affil{Astrophysics Research Centre, School of Mathematics and Physics, Queen's University, Belfast, BT7 1NN, U.K.}
\email{m.cordiner@qub.ac.uk}

\begin{abstract}
A new chemical model is presented for the carbon-rich circumstellar envelope of the AGB star IRC+10216. The model includes shells of matter with densities that are enhanced relative to the surrounding circumstellar medium. The chemical model uses an updated reaction network including reactions from the RATE06 database and a more detailed anion chemistry. In particular, new mechanisms are considered for the formation of CN$^-$, C$_3$N$^-$ and C$_2$H$^-$, and for the reactions of hydrocarbon anions with atomic nitrogen and with the most abundant cations in the circumstellar envelope.  New reactions involving H$^-$ are included which result in the production of significant amounts of C$_2$H$^-$ and CN$^-$ in the inner envelope. The calculated radial molecular abundance profiles for the hydrocarbons C$_2$H, C$_4$H and C$_6$H and the cyanopolyynes HC$_3$N and HC$_5$N show narrow peaks which are in better agreement with observations than previous models.  Thus, the narrow rings observed in molecular microwave emission surrounding IRC+10216 are interpreted as arising in regions of the envelope where the gas and dust densities are greater than the surrounding circumstellar medium. Our models show that CN$^-$ and C$_2$H$^-$ may be detectable in IRC+10216 despite the very low theorised radiative electron attachment rates of their parent neutral species. We also show that magnesium isocyanide (MgNC) can be formed in the outer envelope through radiative association involving Mg$^+$ and the cyanopolyyne species.

\end{abstract}


\section{Introduction}

The carbon-rich AGB star IRC+10216 is among the richest known sources of molecules in the sky.  The star is nearing the end of its life and is rapidly losing mass to the interstellar medium. Over 60 species have been detected in the expanding circumstellar envelope (CSE), which extends more than 10,000 AU from the star. The outflow from the central star is carbon-rich (with [C]/[O]$>$1), and contains abundant stable molecules such as C$_2$H$_2$, HCN and CH$_4$, which are formed in the hot inner regions near the stellar surface \citep{laf82}. As the density falls with radius, interstellar UV photons penetrate the CSE and cause the photolysis of these `parent' molecules.  This results in the production of reactive species that subsequently participate in a complex gas-phase chemistry to produce new `daughter' molecules in the outer envelope. Chemical models of this environment \citep[see][for example]{mil00}, are able to reproduce with good accuracy the observed column densities of outer-envelope species such as the carbon chains C$_n$H (for $n=2, 4, 6, 8$). However, it has been suggested that such chemical models are inconsistent with observed C$_n$H rotational emission maps \citep[see for example][]{gue99}. The observations show C$_n$H emission maxima at around the same radius (for different $n$), whereas the results of chemical models show spatially separated peak abundances. The modeled molecular radial distributions are also broader than observed. Using optical imaging, \citet[][]{mau00} detected multiple dust shells in the envelope of IRC+10216, spaced at regular intervals of $\sim5''$ to $20''$.  These shells were interpreted as arising as a result of modulation of the stellar mass-loss rate, perhaps due to the presence of a binary stellar companion. \citet{bro03} included density-enhanced circumstellar dust shells in their chemical model for IRC+10216 in an attempt to make the model more physically realistic and to address the discrepancies between observed molecular emission maps and the results of previous chemical models. The addition of dust shells modifies the radiation field inside the CSE and results in modeled molecular abundance profiles that are in better agreement with the observations. 

\begin{figure}
\centering
\epsfig{file=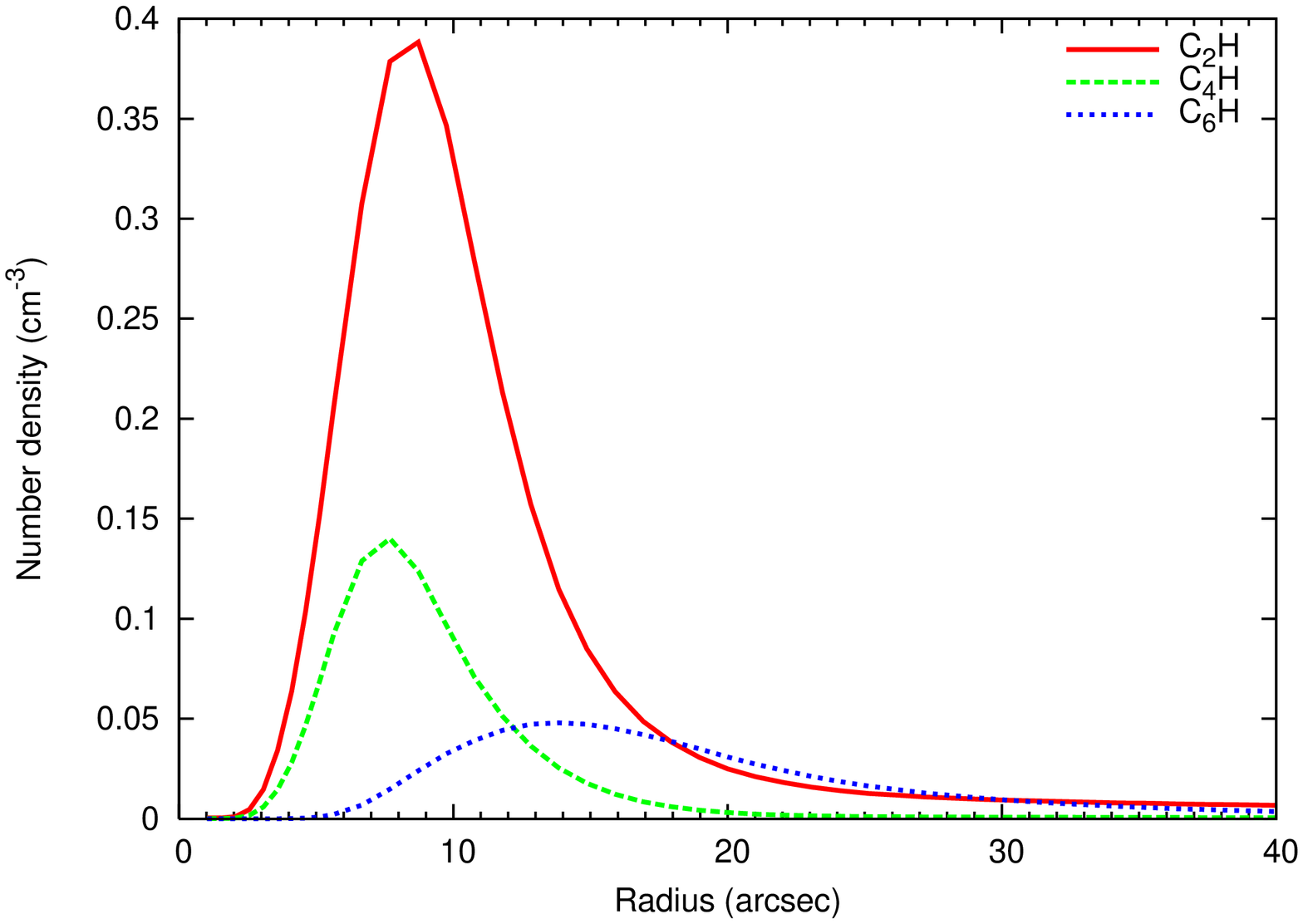, width=0.7\columnwidth, angle=270}
\epsfig{file=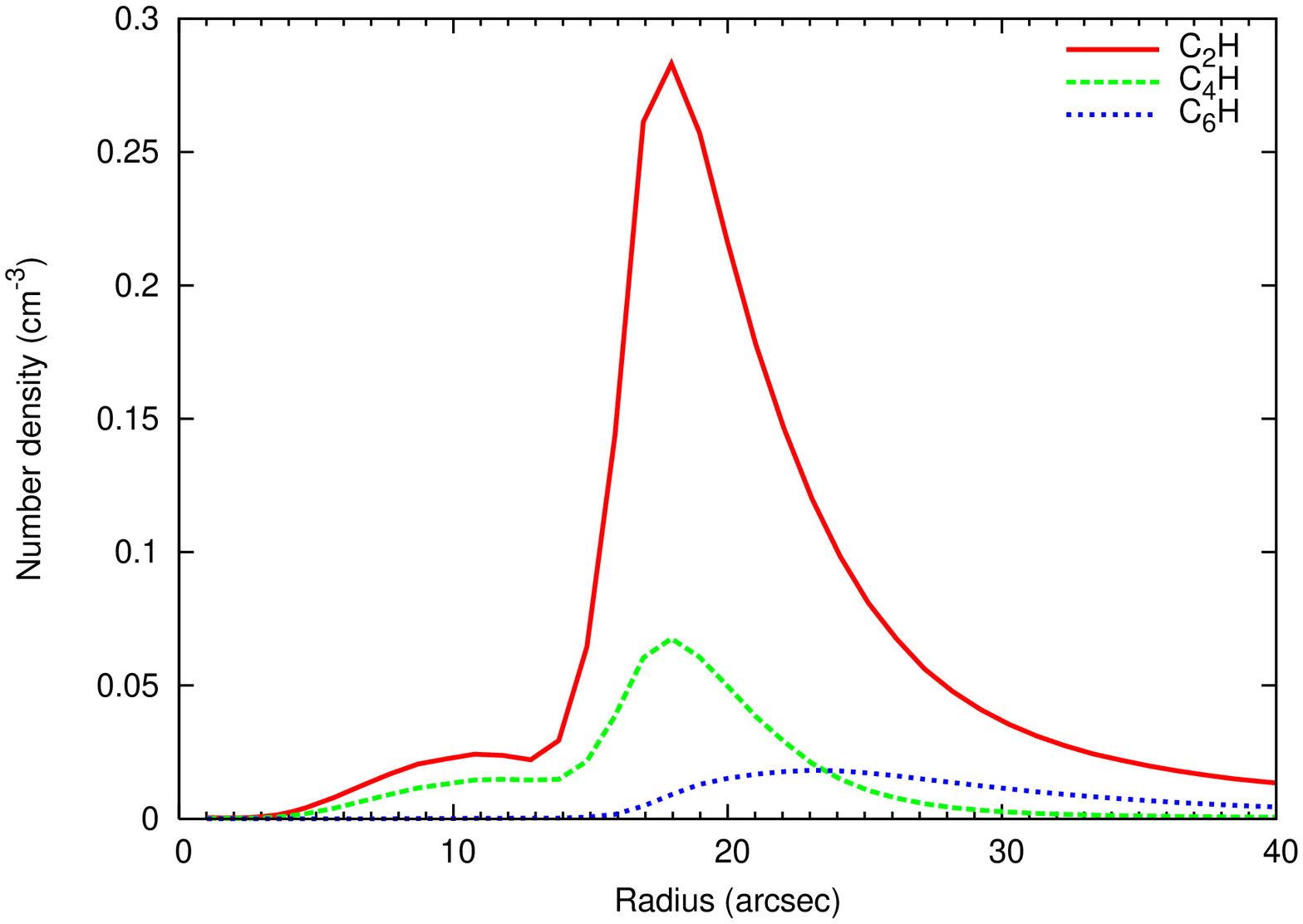, width=0.7\columnwidth, angle=270}
\caption{Radial molecular abundance profiles for C$_2$H, C$_4$H and C$_6$H in IRC+10216, calculated in a model with no density-enhanced shells (top), and with the density-enhanced dust shells of \citet{bro03} (bottom).  The distance to IRC+10216 has been assumed to be 130 pc.  Details of the chemical model used to calculate these profiles are given in \pref{sec:model}.}
\label{fig:no_gas_shells}
\end{figure}

\pref{fig:no_gas_shells} shows the effect of the \citet{bro03} density-enhanced dust shells on the modeled C$_2$H, C$_4$H and C$_6$H radial profiles in IRC+10216.  In the presence of dust shells, the molecular abundance peaks move outwards and slightly closer together.  The C$_2$H profile FWHM narrows from $7''$ to $6''$ and moves from $8''$ to $18''$. According to \citet{gue99}, the molecular emission maps from C$_2$H, C$_4$H and C$_6$H all peak in a narrow circumstellar ring about $2''$ wide at a radius about $15''$ from the star. It is therefore clear that although the radial distributions of the species in \citet{bro03}'s model represent a better match than previous models, they are still too broad to fully explain the observed maps.  In addition, C$_6$H peaks at a larger radius than C$_2$H and C$_4$H, which is inconsistent with the observations.

\citet{din08} mapped the molecular shells of HC$_3$N and HC$_5$N in IRC+10216 at unprecedented angular resolution. The shells were found to be clumpy, co-spatial with each other and with a structure closely matching the distribution of dust shells observed by \citet{mau00}. It was concluded that the circumstellar cyanopolyyne gas density distribution matches that of the dust and therefore that the gas and dust are coupled.  Shell-like structures $\sim2''$ wide are present in the maps of \citet{din08}, which are inconsistent with the $\sim10''$ wide abundance profiles calculated in the model of \citet{bro03}.

In this article, a new chemical model for IRC+10216 is presented which builds on previous models and includes density-enhanced shells of gas in addition to the dust shells of \citet{bro03}. Thus, the new model assumes kinematical coupling between the gas and dust in the CSE.  The aim is to model the chemistry of the narrow shell-like structures observed in the IRC+10216 outflow by \citet{din08} and \citet{gue99} to test the idea that coupling between the gas and dust plays an important role in determining the morphology of the molecular distributions.

The molecular anions C$_4$H$^-$, C$_6$H$^-$, C$_8$H$^-$ and C$_3$N$^-$ have been recently detected in the envelope of IRC+10216 \citep[for a review of the astrophysical anion detections reported so far, see][]{her08}. Following these discoveries we also expand upon the anion chemistry in IRC+10216 studied by \citet{mil00,mil07}.  In particular, the C$_n$N$^-$ chemistry is reconsidered using the reaction rates between N-atoms and carbon-chain anions published by \citet{eic07}. The possibility that magnesium isocyanide (MgNC) is produced in the outer CSE by gas-phase chemistry is also examined.

\section{The chemical model}
\label{sec:model}

\subsection{Physics}

\begin{figure}
\centering
\epsfig{file=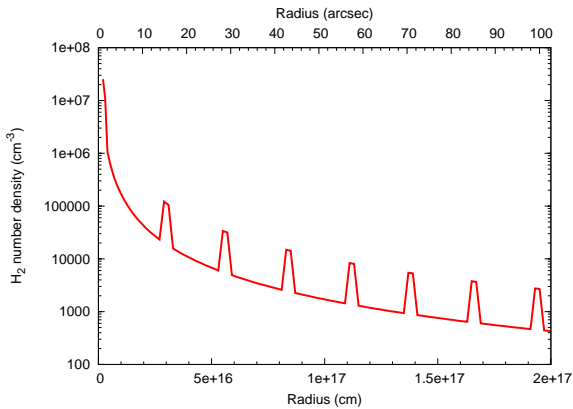, width=0.7\columnwidth, angle=270}
\caption{Radial H$_2$ density profile used in the model.}
\label{fig:density}
\end{figure}

The new chemical model for IRC+10216 is based on the model of \citet{mil00}. The underlying density distribution in the CSE is derived using a mass-loss rate of 1.6$\times10^{-5}$ M$_{\sun}$\,yr$^{-1}$ \citep[after][]{men01}, and assuming that the matter expands in a spherically symmetric outflow with a velocity of 14 \kms. The resulting gas number density profile $n(r)$ falls as $1/r^2$, added to which are a series of step-like density enhancements of the form $\beta n(r)$. Based on the dust shell parameters deduced from scattered light observations by \citet{mau00}, each shell is $2''$ thick with an inter-shell spacing of $12''$, which corresponds to a timescale of approximately 530 years between successive episodes of enhanced mass-loss. The parameter $\beta$ is set to 5 for all shells in the present model (see \pref{sec:evol}). To convert between physical length units and angular distances on the sky, the distance to IRC+10216 is taken to be 130 pc \citep[after][]{men01}. The chosen shell parameters are not intended to provide an accurate representation of the dust shell structure observed in IRC+10216 (for which a 3-dimensional, time-dependent model of the gas and dust in the envelope would be required), but to permit the study of the general effects on the chemical model of the addition of density-enhanced gas and dust shells. The radial H$_2$ gas density distribution used in the model is shown in \pref{fig:density}. The H$_2$ is assumed to be completely self-shielded in the regions of interest in the CSE so that $n_{H_2}(r)\approx n(r)$.

The adopted temperature profile is based on an empirical fit to the gas kinetic temperature profile derived by \citet{cro97}, and takes the form
{\scriptsize
\be
T(r)={\rm max}\left[\left(\frac{2.81\times10^{15}}{r}\right)^{4.7} + \left(\frac{3.34\times10^{17}}{r}\right)^{1.05},10\right]
\ee
}
with a lower limit of 10 K to prevent the temperature becoming unrealistically low in the outer envelope. Parent species (with abundances shown in \pref{tab:parents}) are injected into the model at the inner radius of $r_i=10^{15}$ cm, where the gas density is $8.6\times10^7$ cm$^{-3}$.  At this radius the gas kinetic temperature is 575 K and the interstellar radiation field is attenuated by an effective radial extinction of 47 magnitudes in the $V$ band.  Initial abundances are the same as in \citet{mil00}, with the addition of Mg (see \pref{sec:mgnc}), and with C$_2$H$_2$ and HCN abundances taken from the recent measurements by \citet{fon08}.

\begin{deluxetable}{lc}
\tablecolumns{2}
\tablewidth{0pt}
\tablecaption{Initial fractional abundances of parent species relative to H$_2$ \label{tab:parents}}
\tablehead{
\colhead{Species} & \colhead{Initial abundance}} 
\startdata
He&1.5$\times10^{-1}$\\
C$_2$H$_2$&8.0$\times10^{-5}$\\
CH$_4$&2.0$\times10^{-6}$\\
H$_2$S&1.0$\times10^{-6}$\\
HCN&2.0$\times10^{-5}$\\
NH$_3$&2.0$\times10^{-6}$\\
CO&6.0$\times10^{-4}$\\
CS&4.0$\times10^{-6}$\\
N$_2$&2.0$\times10^{-4}$\\
Mg&1.0$\times10^{-5}$\\
\enddata
\end{deluxetable}

The standard interstellar radiation field \citep{dra78} is assumed to impinge on the outside of the circumstellar envelope from all directions.  The extinction is calculated for the underlying $1/r^2$ density distribution of the CSE using the approach of \citet{jur81}. Additional terms are added to the extinction due to the contributions of the density-enhanced shells \citep[for further details of the extinction calculation in the presence of dust shells, see][]{bro03}.

\subsection{Chemistry}
\label{sec:chem}

The chemical reaction network is based on that used by \citet{pet03} and \citet{mil07}. The reaction rates have been updated to be consistent with those in the (dipole-enhanced) RATE06 database \citep{wood07}. Additional new reactions from RATE06 for those species in the \citet{pet03} model have also been added. The following reactions were deleted from the reaction network in order to increase computational speed, with negligible effect on the chemistry of the species of interest in this study: those with activation energies greater than 300 K, those reactions involving H$_3$O$^+$ as a reagent (except for H$_3$O$^+$ + $e^-$), those with H$_2^+$ as a reagent (except for H$_2^+$ + $e^-$ and H$_2^+$ + H$_2$), and any reactions involving C$^-$, S$^-$, CO$^+$, HOC$^+$, C$_2$H$_6^+$, and C$_2$H$_5$.

Carbon-chain species, hydrocarbons, cyanopolyynes and their associated anions and cations are of principal importance for the chemistry of this study. The following species are among those included in the chemical model: carbon chains C$_n$ ($n=1-23$), C$_n^+$ ($n=1-23$), C$_n^-$ ($n=3-23$); C$_{2n-1}^{(+)}$S ($n=1-3$); HC$_{2n-1}$S ($n=1-3$); hydrocarbons C$_n$H$^{(+/-)}$ ($n=2-23$), C$_n$H$_2^{(+)}$ ($n=1-23$), C$_n$H$_3$ ($n=1-4$), C$_n$H$_3^{+}$ ($n=1-23$), C$_n$H$_4$ ($n=1-3$), C$_n$H$_4^{+}$ ($n=1-9$), C$_n$H$_5^{+}$ ($n=2-9$); and cyanopolyynes C$_{2n-1}$N$^{(+/-)}$ ($n=1-11$), HC$_{2n-1}$N ($n=1-12$), HC$_{2n-1}$N$^+$ ($n=1-11$) H$_2$C$_{2n-1}$N$^+$, ($n=1-11$), H$_3$C$_{2n-1}$N$^+$ ($n=2-5$). 

Due to the current interest in molecular anions in IRC+10216 \citep[\eg][]{mil07,rem07,tha08,cor08}, the anion chemistry has been extended to include C$_n$H$^-$ down to $n=2$, utilizing the radiative electron attachment rates from \citet{her08}. The following additional CN$^-$ and C$_2$H$^-$ formation reactions have been included:
\be
{\rm HCN} + {\rm H^-} \longrightarrow {\rm CN^-} + {\rm H_2}
\label{eq:hcnh-}
\ee
\be
{\rm C_2H_2} + {\rm H^-} \longrightarrow {\rm C_2H^-} + {\rm H_2}
\label{eq:c2h2h-}
\ee
The rate coefficient used for \pref{eq:hcnh-} ($3.8\times10^{-9}$ cm$^3$\,s$^{-1}$),  is an estimate taken from \citet{pra80}. For \pref{eq:c2h2h-}, the rate coefficient ($4.42\times10^{-9}$ cm$^3$\,s$^{-1}$) was measured experimentally by \citet{mack77}.  These proton-transfer reactions are likely to be important in the model due to the large HCN and C$_2$H$_2$ abundances in the stellar outflow. H$^-$ is produced in the model mainly by the cosmic-ray (CR) dissociation of H$_2$ (H$_2$ + CR~$\longrightarrow$~H$^+$ + H$^-$).  This reaction is slow (with a rate coefficient of $3.9\times10^{-21}$ cm$^3$\,s$^{-1}$; \citealt{pra80}), but provides the main source of H$^-$ in the inner envelope. Interior to the second density-enhanced shell ($r\lesssim2\times10^{16}$ cm), the modeled H$^-$ abundance is about 10$^{-8}$ cm$^{-3}$.  Other  reactions similar to \pref{eq:c2h2h-} were studied by \citet{mack77}, who found that many different molecular anions could recieve a proton from C$_2$H$_2$ at rapid rates ($\sim 10^{-9}$ cm$^3$\,s$^{-1}$), resulting in the production of C$_2$H$^-$. However, it is presently unknown whether reactions occur between C$_2$H$_2$ and the carbon chain anions C$_n^-$ and C$_n$H$^-$. In light of an observational upper limit for $N$(C$_2$H$^-$) in IRC+10216, \citet{cor08} deduced that these reactions probably do not proceed rapidly.

Dominant anion destruction mechanisms are by reaction with H and C$^+$ and by photodetachment.  Photodetachment rates were calculated according to Equation (2) of \citet{mil07}. As a result of its large electron detachment energy, the CN$^-$ photodetachment rate thus calculated is $\sim100$ times less than the value in the RATE06 database.  

Carbon chain anions C$_n^-$ ($n=2-7$), and C$_n$H$^-$ ($n=2, 4, 6$) have been shown to react with atomic nitrogen and result in the formation of products that include the nitrile anions C$_n$N$^-$ ($n=1, 3, 5$) \citep{eic07}. We have included these reactions in the model as part of our investigation into the possible mechanisms for the formation of CN$^-$ and the recently discovered C$_3$N$^-$ \citep{tha08}.  Branching ratios were calculated from \citet{eic07}'s original experimental data by Veronica Bierbaum (private communication). The branching ratio for the associative electron detachment (AED) product channel could not be derived from the experimental data, so we have assumed, arbitrarily, a ratio of 0.5. This assumption constitutes potentially the most significant source of error in these reaction rates.

Mutual neutralisation reactions between anions and cations were shown by \citet{lep88} to have important effects on interstellar chemistry. Thus, reactions of the kind
\be
{\rm X}^+ + {\rm Y}^- \longrightarrow {\rm X} + {\rm Y}
\ee
have been included for the twenty most abundant cations (X$^+$), and for all anions (Y$^-$), in the model, with a rate coefficient of 7.5$\times10^{-8}(T/300)^{-0.5}$ cm$^3$\,s$^{-1}$ \citep[see][for example]{har08}.

The final chemical network contains 426 gas-phase species coupled by 5539 reactions.

\subsection{Evolution of the shells}
\label{sec:evol}

Initially the rate equations are solved as a function of radius \citep[in the same way as][]{mil00}, starting from $r_i=10^{15}$ cm and moving out to $r_f=3\times10^{18}$ cm where photodissociation destroys all the molecules (apart from H$_2$). Then in a separate routine the chemical abundances inside the density-enhanced shells are calculated as a function of radius. The chemical rate equations of a density-enhanced packet lying halfway between a shell's inner and outer radius (at a radius $r_p$ with density $(\beta+1) n(r_p)$), are solved, starting from $r_i$ and moving out to  $2\times10^{17}$ cm where the density is sufficiently low that the shell no longer makes any significant contribution to the total amount of matter in the model. The motion of the density-enhanced shells in the outflow are followed so that they move outward over time, synchronised with the outward radial motion of the dense packet. The radiation field is recalculated for the dense packet at each time-step, taking into account the new positions of all the shells.  The density factor ($\beta$), the inter-shell spacing and the shell thickness are identical for every shell so that the chemical abundances calculated for the dense packet at a given radius represent the abundances in a density-enhanced shell centered at that radius.

\subsection{Model variations}

The effects on the model results of variations in the mass-loss rate, radiation field strength, gas-to-dust ratio, shell thickness, inter-shell spacing, density-enhancement factor and stellar distance have been analysed.  The results presented in this study are for the model that uses the parameters that we believe best match the observational constraints.  Modification of these parameters, particularly those that affect the radiation field strength, can significantly alter the radial abundance profiles calculated by the model.  However, under such circumstances the main conclusions of this study remain the same.  The location of the $15''$ density-enhanced shell has been fixed in order to best reproduce the observational data, which may be considered a contrivance of the model.  However, in Figure 8 of \citet{mau00}, this can be identified as the radius at which the first distinct dust shells occur.

To permit comparison between this model and previous models published in the literature (which do not include density-enhanced gas and dust shells), we have also run the model without any density-enhanced shells (\ie\ $\beta=0$) and also with the density-enhanced dust shells of \citep{bro03} to produce the data shown in \pref{fig:no_gas_shells}.

\subsection{Molecular excitation}

To facilitate comparison of the model results with observed maps of molecular microwave emission line flux, the rotational excitation of some of the molecules of interest has been calculated as a function of radius using a modified version of the mmline computer code \citep[described by][]{jus94}.  The central stellar radius was taken to be 3.1$\times10^{13}$ cm and the temperature 2650 K \citep[derived from][]{men01}.  The dust opacity was taken from Figure 6 of \citet{men01}.   Due to the lack of published collisional excitation rates and vibrational transition strengths, only a rough estimate of the rotational excitation is possible for most molecules. For C$_2$H and C$_2$H$^-$ the HCO$^+$ rates from \citet{flo99} are used, and for C$_4$H, C$_4$H$^-$, C$_6$H and C$_6$H$^-$ the HC$_3$N rates from \citet{gre78} are used. For C$_6$H and C$_6$H$^-$ the rates have been extrapolated up to $J=31$. The collisional rates used are for closed electronic-shell species so the spectroscopic structure of the corresponding (closed electronic-shell) anions has been used for the calculations of the excitation of the open-shell C$_2$H, C$_4$H and C$_6$H radicals. This approximation is reasonable because the structure of these hydrocarbon anions and neutrals are very similar. Only the $^2\Pi_{1/2}$ states have been considered for the neutral hydrocarbons; the population of the $^2\Pi_{3/2}$ states are not expected to significantly affect the relative populations of the states of interest here. Rotational Einstein $A$ coefficients were calculated using dipole-moments from \citet{woo95} and \citet{bla01} for the neutrals and the anions, respectively. IR pumping of rotational levels has been calculated through consideration of the radiative excitation of a single vibrational state $\sim10$ $\mu$m above the ground state \citep[\emph{cf}.][]{bie84}.  C$_2$H has been calculated to have a strong vibrational transition ($A=0.6$ s$^{-1}$) at 12.5 $\mu$m \citep{tar04}, which we assume to also occur in C$_2$H$^-$. The vibrational spectra of C$_4$H, C$_6$H and their associated anions are less well known. For these species a transition has been assumed to occur at 12.5 $\mu$m with $A=1$ s$^{-1}$.  IR pumping has a significant effect on the molecular excitation, but changes in the vibrational transition wavelengths and Einstein $A$ coefficients by up to an order of magnitude do not significantly affect the results of the present study. To assess the impact of errors in the collisional excitation rates on the calculated molecular emission profiles, the rates were varied by an order of magnitude either way.  The overall features of the emission profiles remained the same. 

\section{Results}

\subsection{Molecular radial abundance and intensity distributions}

\begin{figure}
\centering
\epsfig{file=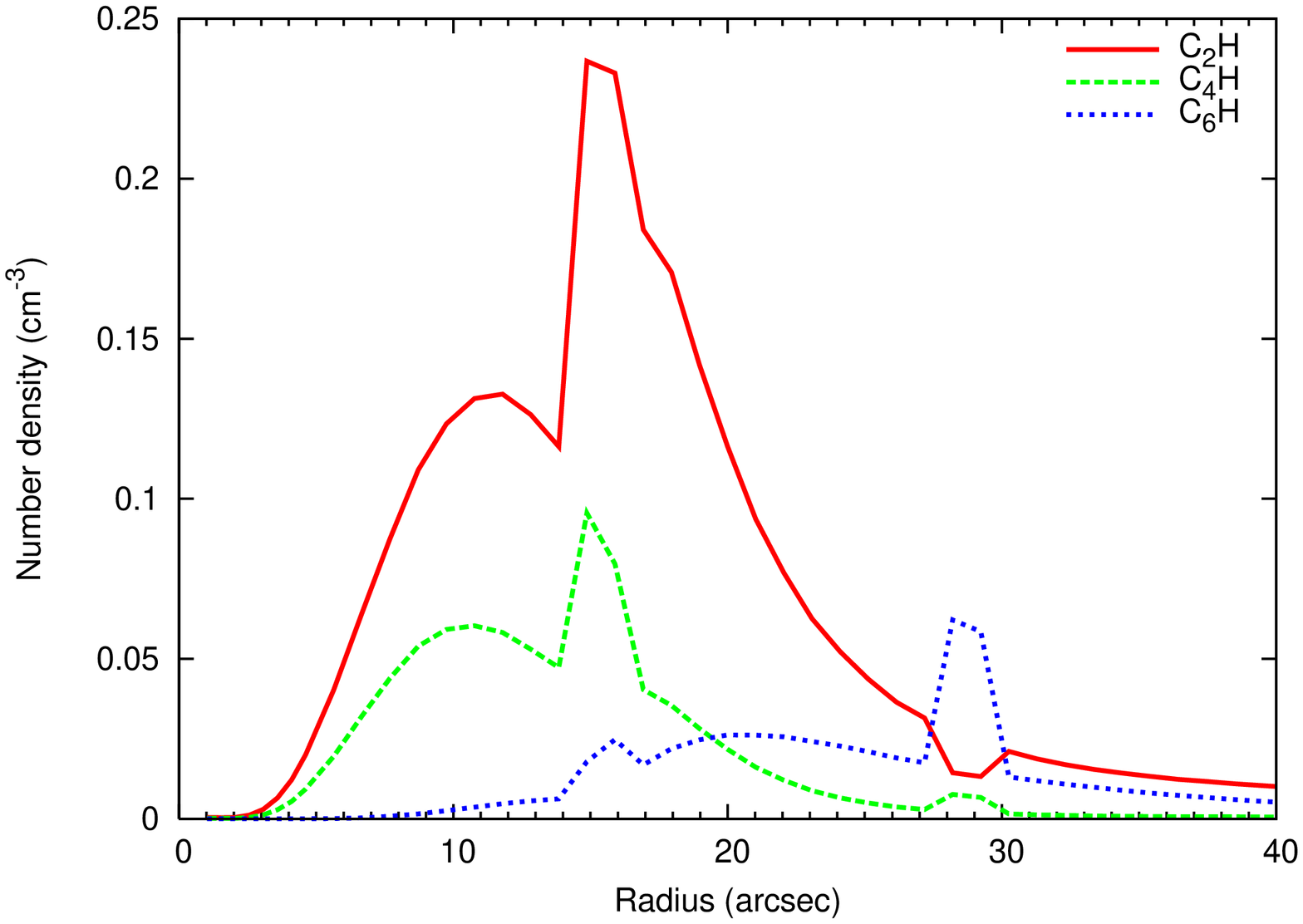, width=0.7\columnwidth, angle=270}
\epsfig{file=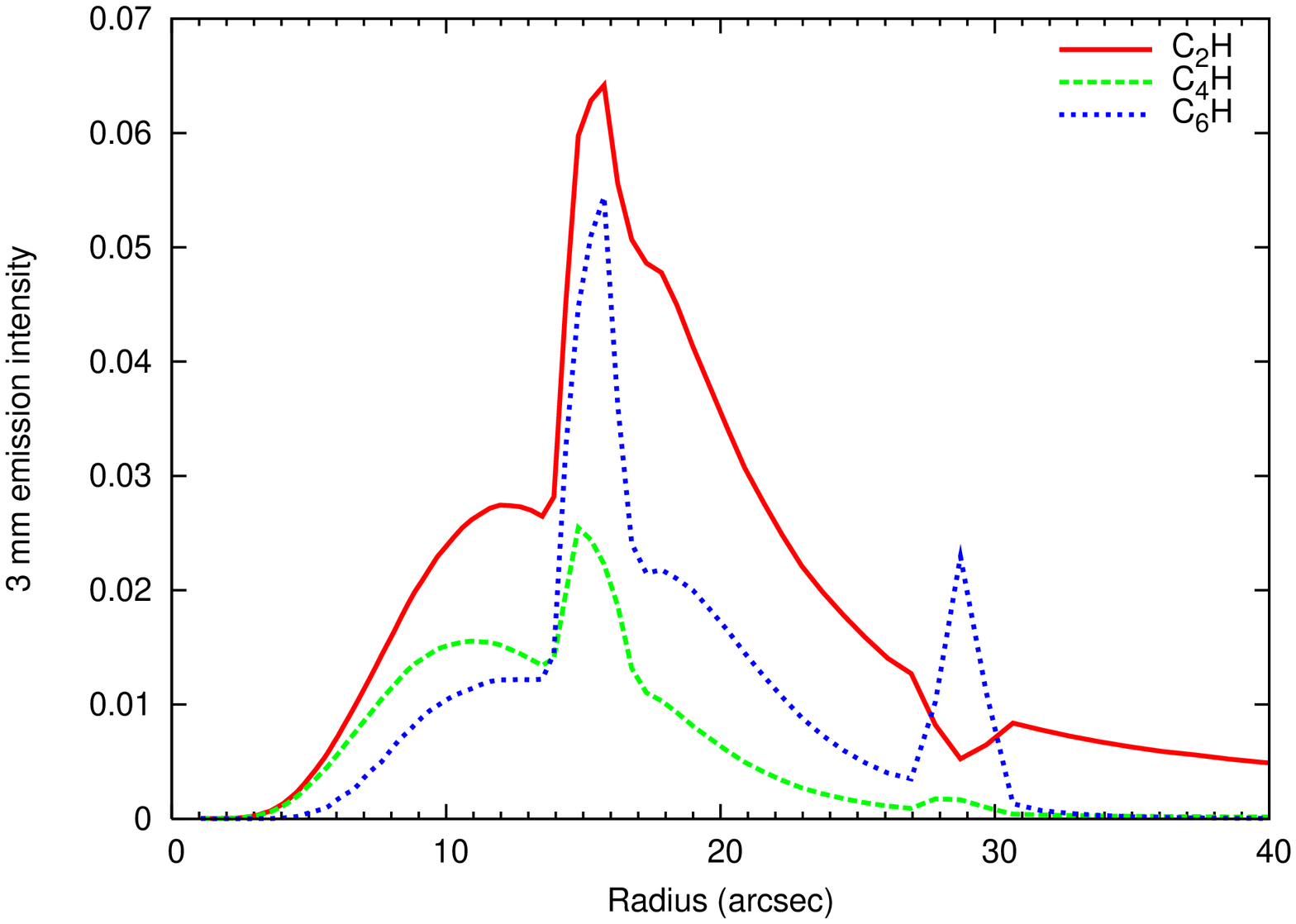, width=0.7\columnwidth, angle=270}
\caption{C$_2$H, C$_4$H and C$_6$H modeled radial abundance profiles (top) and 3 mm emission intensity profiles (multiplied by constant scaling factors for display) (bottom).}
\label{fig:CnH}
\end{figure}

\begin{figure}
\centering
\epsfig{file=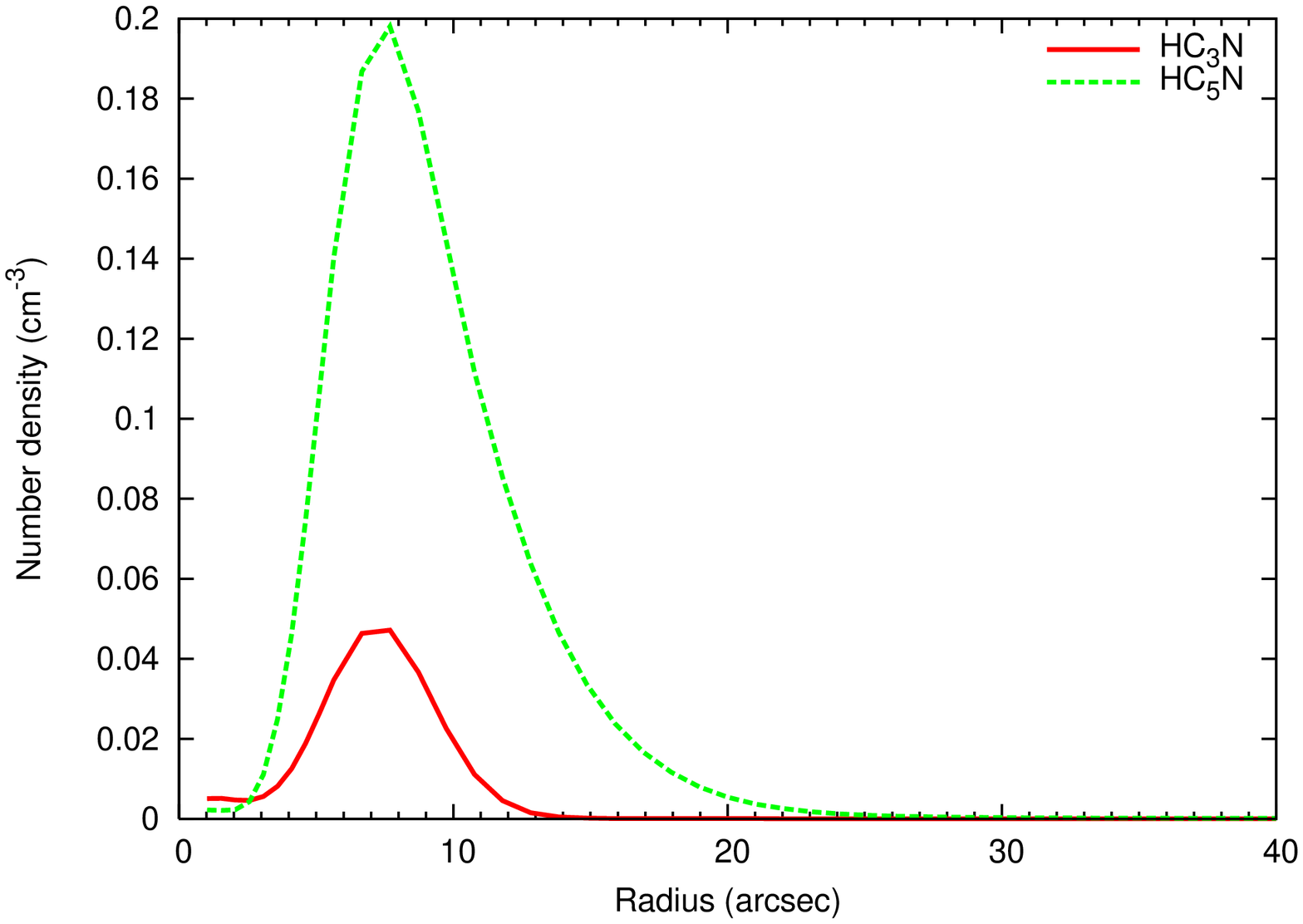, width=0.7\columnwidth, angle=270}
\epsfig{file=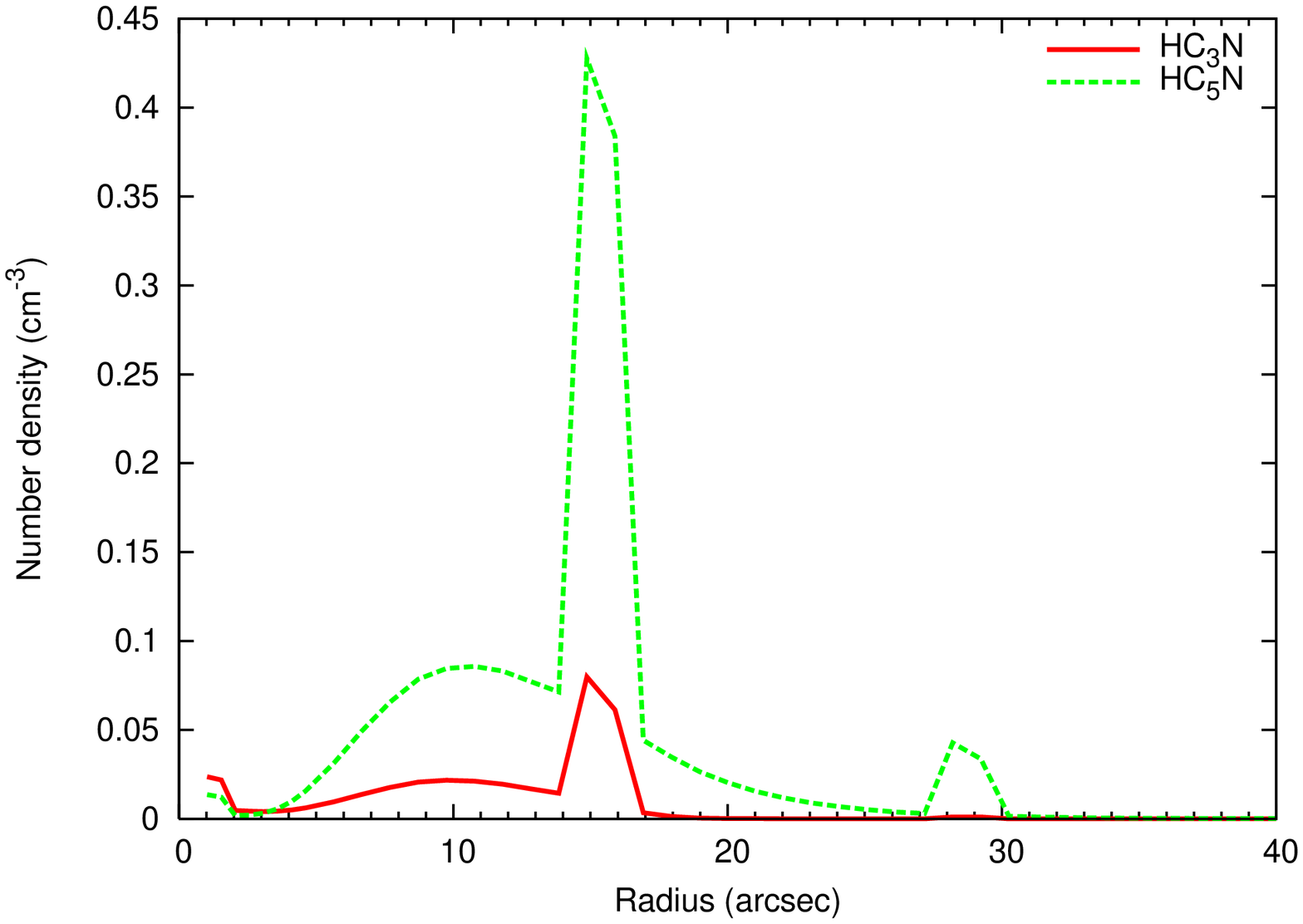, width=0.7\columnwidth, angle=270}
\epsfig{file=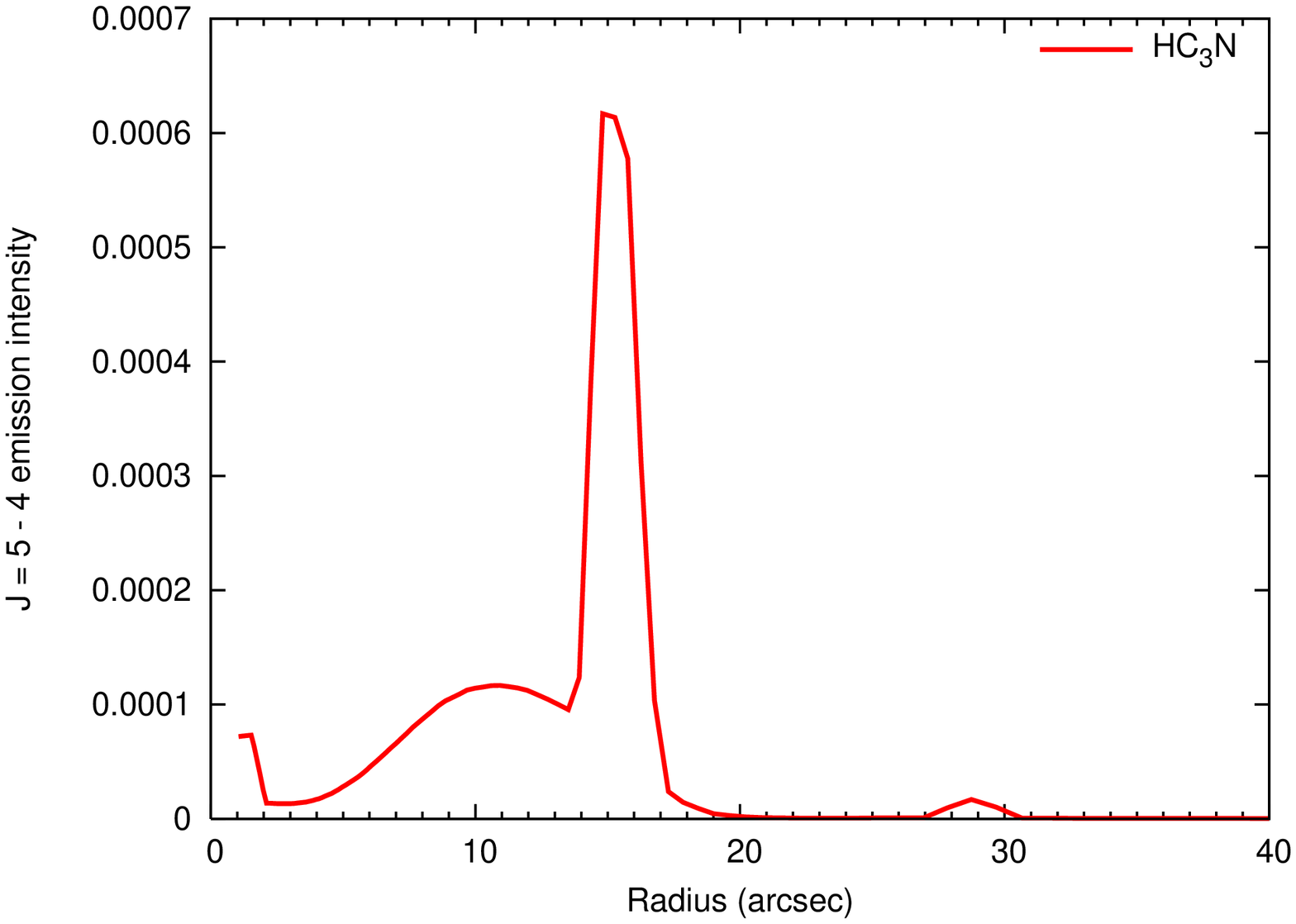, width=0.7\columnwidth, angle=270}
\caption{Modeled radial abundance profiles for HC$_3$N and HC$_5$N in the absence of density-enhanced shells (top) and with density-enhanced shells (middle).  The bottom panel shows the modeled HC$_3$N $J=5-4$ intensity profile calculated in the presence of density-enhanced shells.}
\label{fig:cyano}
\end{figure}

\begin{figure}
\centering
\epsfig{file=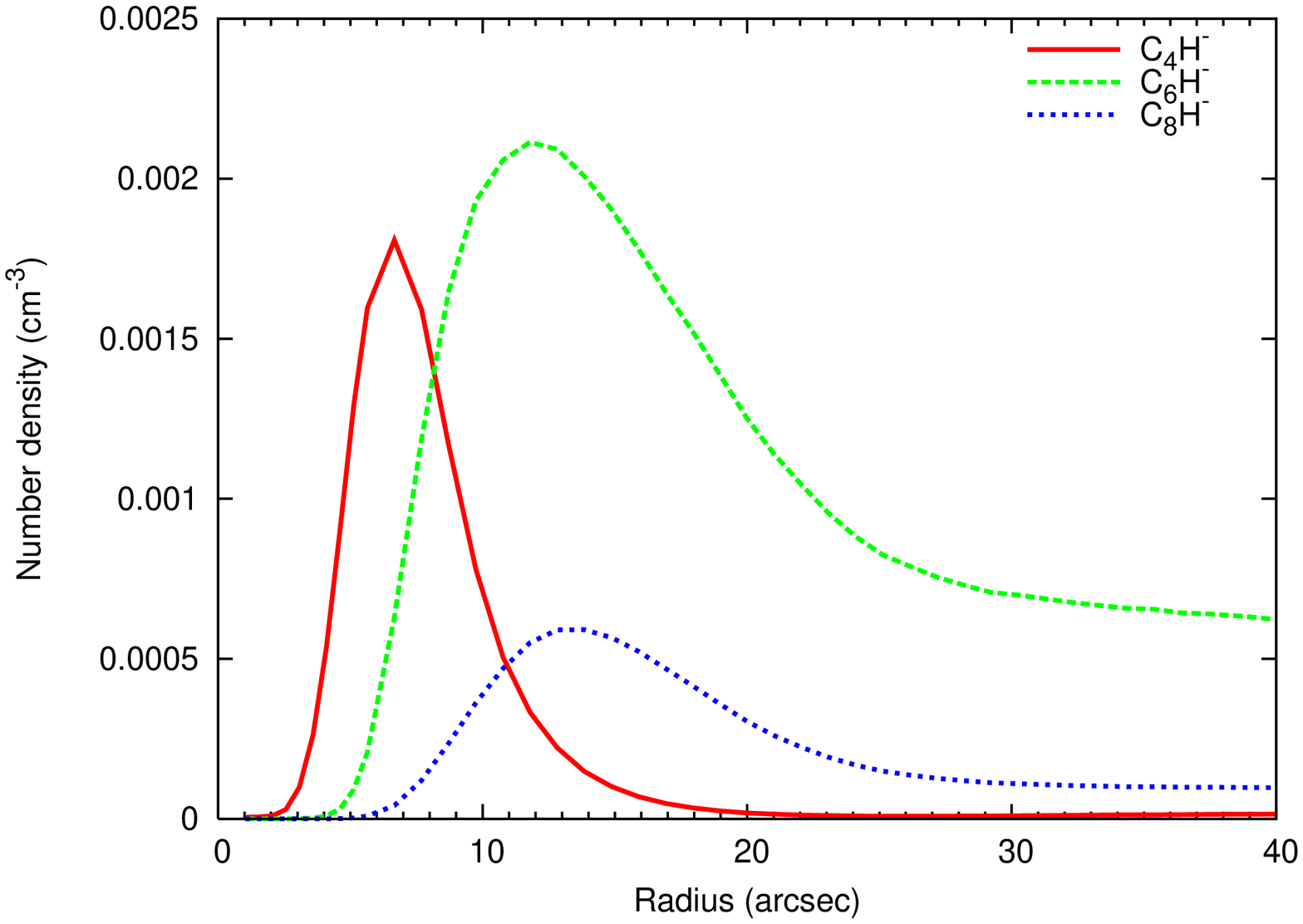, width=0.7\columnwidth, angle=270}
\epsfig{file=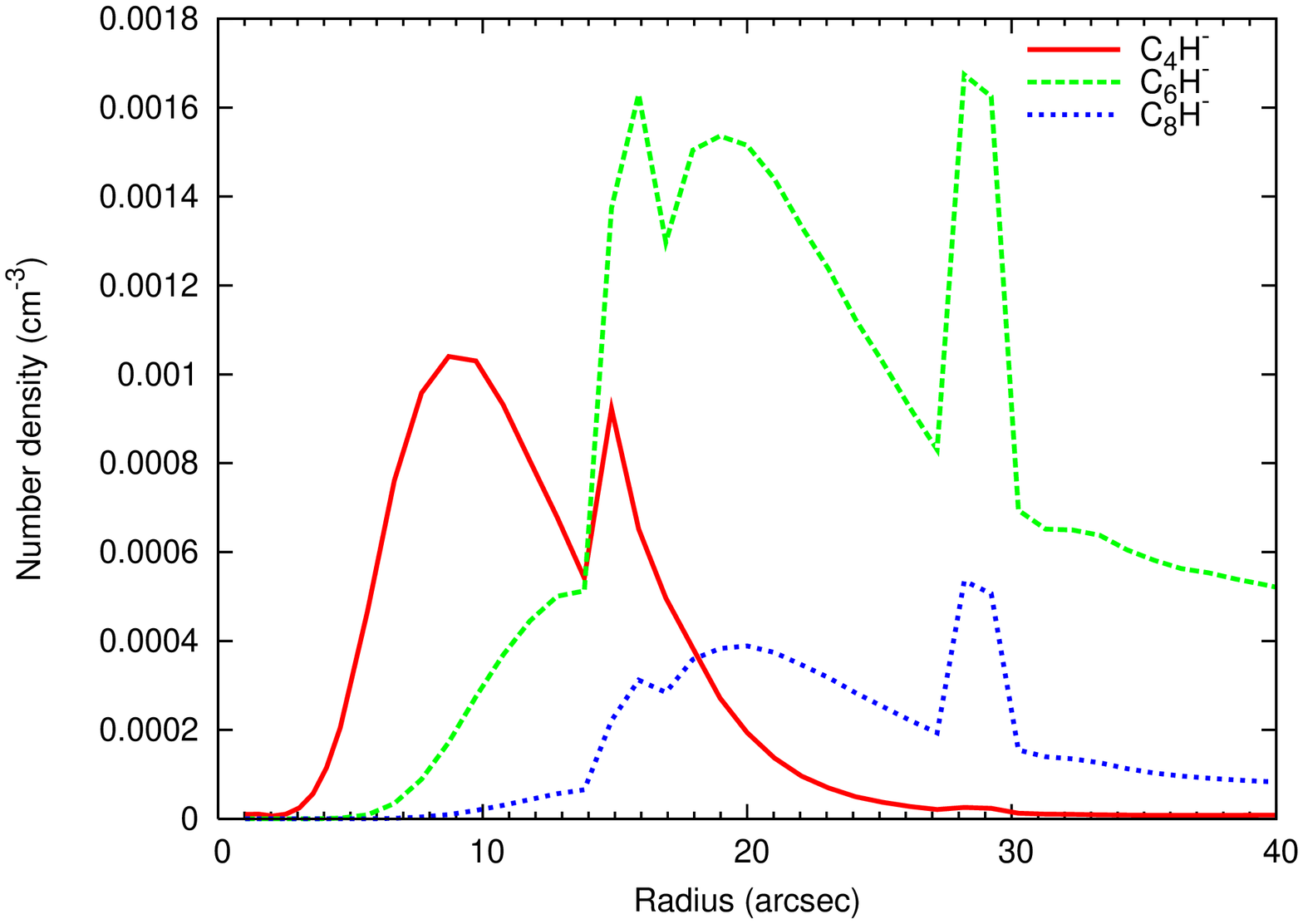, width=0.7\columnwidth, angle=270}
\epsfig{file=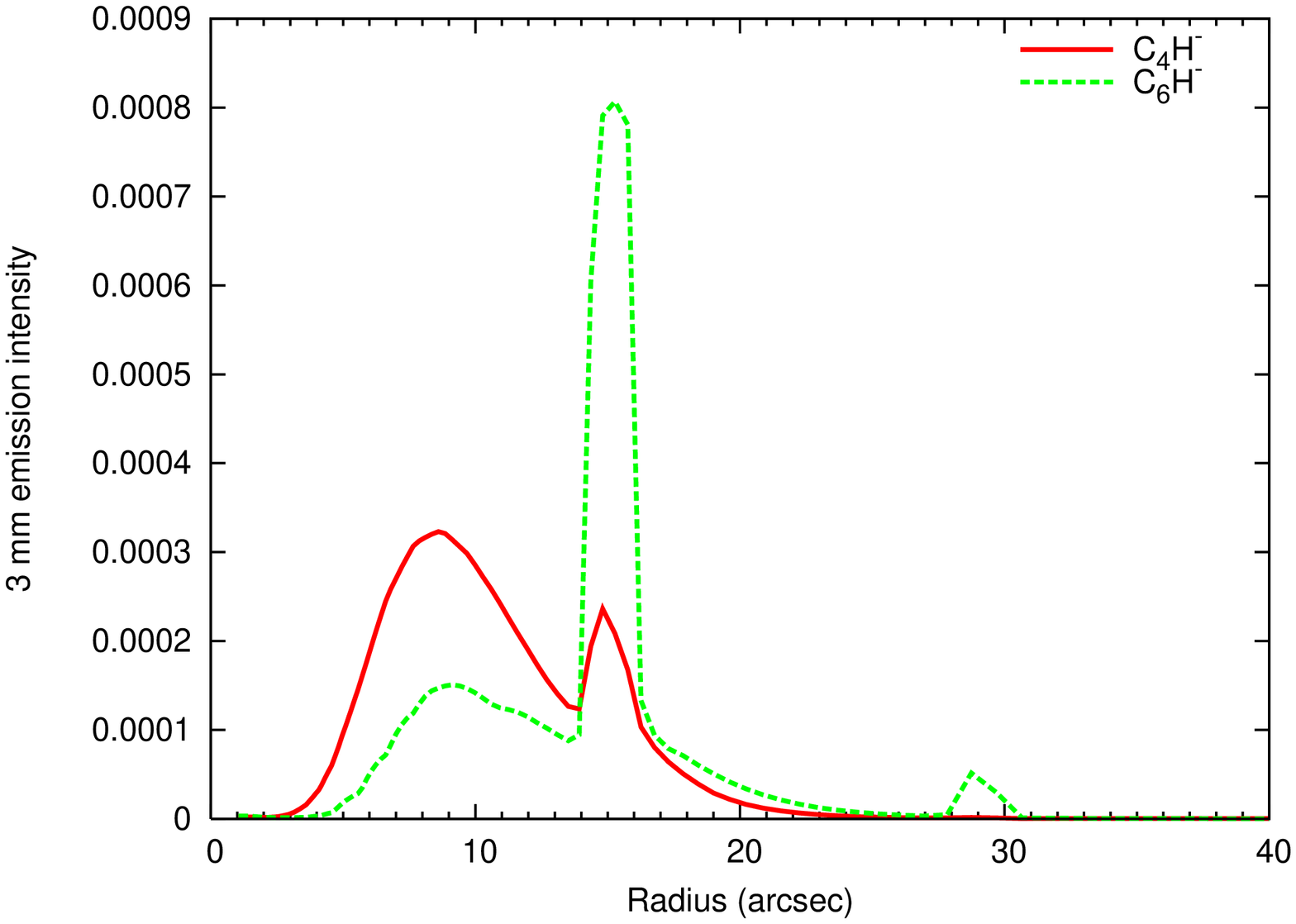, width=0.7\columnwidth, angle=270}
\caption{Modeled radial abundance profiles for the hydrocarbon anions C$_4$H$^-$, C$_6$H$^-$ and C$_8$H$^-$ in the absence of density-enhanced shells (top) and with density-enhanced shells (middle). C$_4$H$^-$ and C$_6$H$^-$ 3 mm emission intensity profiles are shown in the bottom panel.}
\label{fig:anions}
\end{figure}

The upper panel of \pref{fig:CnH} shows the calculated radial abundance profiles for the hydrocarbons C$_n$H ($n=2,4,6$).  Comparison with \pref{fig:no_gas_shells} shows the impact of the density-enhanced shells on these species.  A prominent effect of the shells is to raise of the abundances relative to the surrounding CSE due to the increased gas density. The photon-induced hydrocarbon chemistry is suppressed very near to the star by the dust shell at a radius of $r=1''$ which provides additional shielding of C$_2$H$_2$ from photodissociation.  The abundances all reach a peak in the shell at $r=15''$. However, this is not where C$_6$H reaches its greatest abundance, which occurs in the third density-enhanced shell at $r=29''$. As shown in the middle panel of \pref{fig:cyano}, HC$_3$N and HC$_5$N both reach their greatest abundances in the $15''$ shell.  Cosmic-ray-induced chemistry also results in the synthesis of a significant amount of HC$_3$N in the innermost ($r=1''$) shell (through the reaction HCN + CRPHOT $\longrightarrow$ CN + H, followed by CN + C$_2$H$_2 \longrightarrow$ HC$_3$N + H).  The anions C$_4$H$^-$, C$_6$H$^-$ and C$_8$H$^-$ (shown in \pref{fig:anions}), do not reach their maximum abundances in the $15''$ shell; C$_4$H$^-$ peaks at a similar radius to the model with no shells whereas C$_6$H$^-$ and C$_8$H$^-$ peak in the $29''$ shell.

Molecular abundances tend to peak within the density-enhanced shells because the increased density raises the abundances of chemical reagents which drives the chemistry at a faster rate.  It is not always the case, however, that this raises the abundances of daughter species, as can be seen in \pref{fig:CnH} where the C$_2$H abundance is reduced in the $r=29''$ shell due to the increased densities of atoms and ions (including C, C$^+$ and N), that it reacts with.

A comparison of the abundance profiles in the models with and without density-enhanced shells shows that, in general, the increased shielding of the CSE from interstellar UV by the dust shells (which inhibits the photochemistry), causes photodissociation of parent species to be less efficient in the inner regions and causes the daughter abundances to rise more slowly with radius, moving the profile maxima outwards.

The calculated emission intensities for C$_2$H, C$_4$H, C$_6$H, C$_6$H$^-$ and HC$_3$N (shown in the lower panels of Figures \ref{fig:CnH}, \ref{fig:cyano} and \ref{fig:anions}, respectively), are greatest within the $r=15''$ density-enhanced shell.  C$_4$H$^-$ reaches maximum intensity near $r=8''$.  The differences between the abundance profiles and the emission profiles for C$_6$H and C$_6$H$^-$ are particularly notable. Because the 3 mm emission from these species originates from a high rotational level (around $J=30$), the strength of the emission is highly dependent on the rate of collisional excitation, and therefore the density. Thus, inside the density-enhanced shells the lower $J$ levels tend to become depopulated in favour of the higher levels.

\begin{figure}
\centering
\epsfig{file=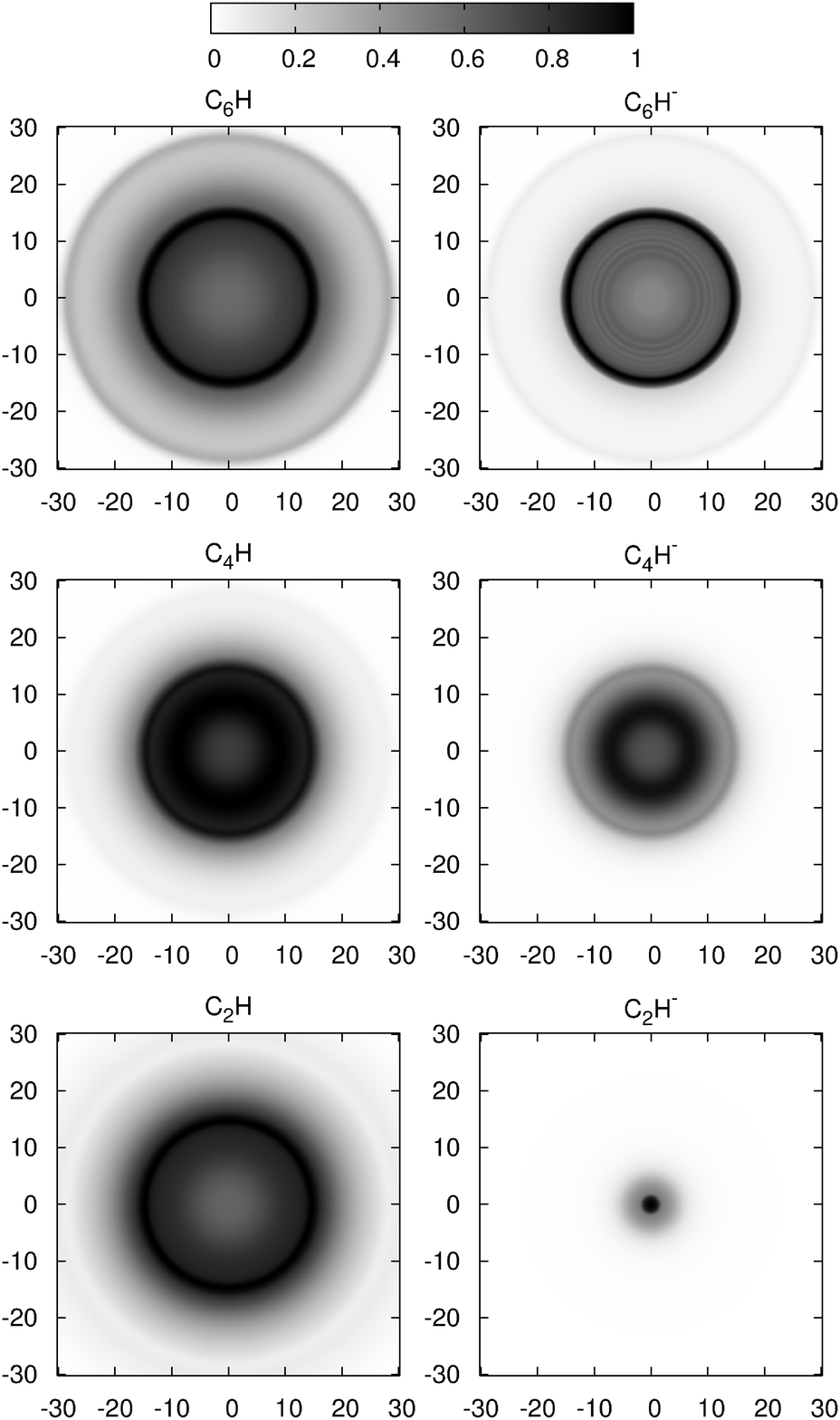, width=0.95\columnwidth}
\caption{Emission maps for hydrocarbons and their anions, calculated from the modeled radial intensity profiles assuming spherical symmetry of the CSE. The grey scale is set to one at the respective emission maxima of each plot. Spatial units are arcseconds from the central star.}
\label{fig:maps}
\end{figure}

For C$_2$H, C$_4$H, C$_6$H and their corresponding anions, two-dimensional emission maps have been constructed from the radial emission profiles by integration along (pencil-beam) lines of sight through the CSE.  This calculation assumes spherical symmetry of the CSE and therefore cannot show any of the azimuthal structure in observed emission maps. Nevertheless, these maps (shown in \pref{fig:maps}), provide a useful means for comparing the main features in the modeled and observed molecular emission.

C$_2$H and C$_6$H show a strong, narrow emission ring centered on $15''$, in excellent agreement with observed maps of \citet{gue99}. In addition to the observed $15''$ ring, the modeled C$_4$H emission map shows a thick, strong ring at around  $r=8'$.  The patterns for C$_4$H$^-$ and C$_6$H$^-$ are similar to their parent neutrals whereas C$_2$H$^-$ is markedly different: it has a centrally-peaked emission map because it is produced predominantly by the reaction of H$^-$ with C$_2$H$_2$ --- the abundances of which are greatest in the (well shielded) inner CSE ---  rather than by electron attachment to its parent neutral.

\subsection{New anion results}

\begin{deluxetable}{lcccrr}
\tablewidth{0pt}
\tabletypesize{\scriptsize}
\tablecaption{Calculated column densities of negatively-charged species \label{tab:anioncolds}}
\tablehead{\colhead{Species}&\colhead{$N$ (cm$^{-2}$)}&\colhead{Species}&\colhead{$N$ (cm$^{-2}$)}&\colhead{Species}&\colhead{$N$ (cm$^{-2}$)}}
\startdata
C$_3^-$ & 3.9e10  & C$_2$H$^-$ & 5.5e10  &   CN$^-$ & 1.3e12   \\    
C$_4^-$ & 7.5e12  & C$_3$H$^-$ & 1.4e9   &   C$_3$N$^-$ & 9.0e11  \\     
C$_5^-$ & 3.2e13  & C$_4$H$^-$ & 2.5e13  &   C$_5$N$^-$ & 1.7e13  \\   
C$_6^-$ & 5.3e13  & C$_5$H$^-$ & 1.2e13  &   C$_7$N$^-$ & 7.2e12  \\    
C$_7^-$ & 1.6e14  & C$_6$H$^-$ & 9.6e13  &   C$_9$N$^-$ & 3.6e11  \\     
C$_8^-$ & 9.6e13  & C$_7$H$^-$ & 5.2e13  &   C$_{11}$N$^-$ & 3.7e10 \\      
C$_9^-$ & 1.1e14  & C$_8$H$^-$ & 2.0e13  &   C$_{13}$N$^-$ & 2.6e10 \\      
C$_{10}^-$ & 4.0e13 & C$_9$H$^-$ & 1.6e13  &   C$_{15}$N$^-$ & 1.9e10 \\    
C$_{11}^-$ & 3.1e13 & C$_{10}$H$^-$ & 5.2e12 &   C$_{17}$N$^-$ & 1.3e10 \\      
C$_{12}^-$ & 8.3e12 & C$_{11}$H$^-$ & 1.4e13 &   C$_{19}$N$^-$ & 8.5e9  \\     
C$_{13}^-$ & 1.5e13 & C$_{12}$H$^-$ & 3.1e12 &   C$_{21}$N$^-$ & 5.3e9  \\    
C$_{14}^-$ & 3.9e12 & C$_{13}$H$^-$ & 8.4e12 &            &        \\   
C$_{15}^-$ & 4.6e12 & C$_{14}$H$^-$ & 1.4e12 &   CH$_2$CN$^-$ & 2.5e6 \\      
C$_{16}^-$ & 1.8e12 & C$_{15}$H$^-$ & 6.4e12 &              &      \\     
C$_{17}^-$ & 2.5e12 & C$_{16}$H$^-$ & 1.1e12 &   H$^-$ & 3.3e8     \\        
C$_{18}^-$ & 1.1e12 & C$_{17}$H$^-$ & 4.6e12 &         &           \\     
C$_{19}^-$ & 1.5e12 & C$_{18}$H$^-$ & 9.4e11 &   e$^-$ & 4.8e15    \\      
C$_{20}^-$ & 6.9e11 & C$_{19}$H$^-$ & 3.2e12 &         &           \\       
C$_{21}^-$ & 1.0e12 & C$_{20}$H$^-$ & 7.8e11 &         &           \\    
C$_{22}^-$ & 4.2e11 & C$_{21}$H$^-$ & 2.1e12 &         &           \\       
C$_{23}^-$ & 8.1e11 & C$_{22}$H$^-$ & 5.5e11 &         &           \\       
        &        & C$_{23}$H$^-$ & 9.9e12 &         &           \\     

\enddata
\end{deluxetable}

Column densities for all anions included in the chemical model are given in \pref{tab:anioncolds}.

\citet{cor08} identified that the reaction of H$^-$ with HCN (\pref{eq:hcnh-}) could dominate the synthesis of CN$^-$ in IRC+10216. However, the present study shows that reactions between atomic nitrogen and the carbon chain anions may provide a greater source of CN$^-$. Upon inclusion of the reactions N + C$_n^-$ and N + C$_m$H$^-$ into the chemical network (see \pref{sec:chem}), the CN$^-$ column density is raised from $1.3\times10^{10}$ to $1.3\times10^{12}$ cm$^{-2}$.  Due to the fact that C$_7^-$ is the most abundant anion involved in these reactions, the dominant formation reaction for CN$^-$ is N + C$_7^-$~$\longrightarrow$~CN$^-$ + C$_6$.  These results are dependent on the values of the branching ratios assumed for the associative electron detachment reactions.  Even if the AED branching ratio is as large as 0.9 (instead of the assumed value of 0.5), the CN$^-$ column density is calculated to be $2.7\times10^{11}$ cm$^{-2}$ and N + C$_7^-$ is still the dominant reaction for CN$^-$ production.

C$_3$N$^-$ was detected for the first time in IRC+10216 by \citet{tha08}, who report a column density of $1.6\times10^{12}$ cm$^{-2}$. If C$_3$N$^-$ is assumed to be formed only by radiative electron attachment to C$_3$N, the modeled column density is $1.1\times10^{10}$ cm$^{-2}$, with a corresponding anion-to-neutral ratio of 0.02\%. This would imply that the electron attachment rate used \citep[calculated by][]{pet97}, is too small. However, upon inclusion of reactions between nitrogen atoms and anions, the modeled C$_3$N$^-$ column density rises to $9.0\times10^{11}$ cm$^{-2}$ (which corresponds to an anion-to-neutral ratio of 1.3\%), in reasonable agreement with the observed value.  The dominant reactions involved in the production of C$_3$N$^-$ in our present model are N + C$_n^-$ ($n=5,6,7$). If these reactions do indeed dominate the C$_3$N$^-$ synthesis then the rate of radiative electron attachment to C$_3$N calculated by \citet{pet97} may still be accurate. 

Cyanopolyynes C$_n$N (for $n>4$), by analogy with the structurally similar linear hydrocarbons C$_n$H ($n>5$) studied by \citep{her08}, have been assumed to undergo rapid radiative stabilisation upon attachment of a free electron so that the rate of radiative electron attachment used in the model is $1.25\times10^{-7}(T/300)^{-0.5}$ cm$^{-3}$\,s$^{-1}$. Consequently, C$_5$N$^-$ and C$_7$N$^-$ have large modeled column densities of $1.7\times10^{13}$ and $7.2\times10^{12}$  cm$^{-2}$ respectively, which correspond to about 8\% of the abundance of their neutral parents C$_5$N and C$_7$N.  Radiative electron attachment dominates the production of these anions in the model, but it seems plausible that reactions between atomic nitrogen and carbon-chain anions longer than those studied by \citet{eic07} would also result in fragmentation of the carbon chain.  Such fragmentation was observed in the experiments by \citet{eic07} (who studied carbon chain lengths up to only $n=7$ carbon atoms), which gave rise to products with carbon chains with lengths from 1 to $n$.  If the fragmentation of carbon-chain anions with $n>7$ was included in our chemical model, there would be a significant increase in the production rates of the nitrile anions C$_{2n-1}$N$^-$ ($n=1-4$), so that these reactions could potentially dominate the production of C$_5$N$^-$ and C$_7$N$^-$ as well.

\begin{figure}
\centering
\epsfig{file=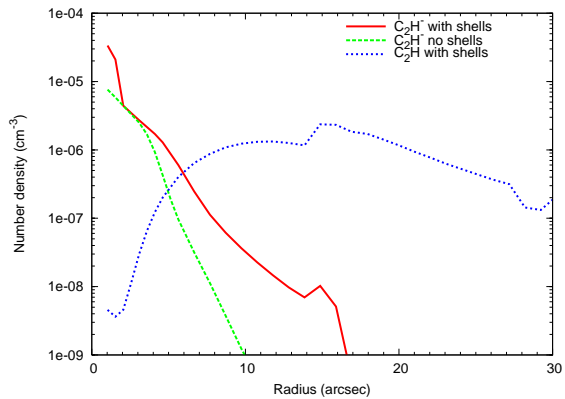, width=0.7\columnwidth, angle=270}
\caption{Modeled C$_2$H$^-$ number density profiles with and without density-enhanced shells. The C$_2$H abundance is also shown, multiplied by $10^{-5}$ for display.}
\label{fig:c2h-}
\end{figure}

Dense shells in the inner envelope have a significant impact on the C$_2$H$^-$ radial abundance profile, as shown in \pref{fig:c2h-}. C$_2$H$^-$ is concentrated in the region of the envelope around the $r=1''$ shell.  The presence of this shell raises the C$_2$H$^-$ column density from $2.3\times10^{10}$ to $5.5\times10^{10}$ cm$^{-2}$. This anion is unusual in the model because its rate of formation by radiative electron attachment is very slow; it is produced predominantly in the inner envelope as a result of \pref{eq:c2h2h-} (H$^-$ + C$_2$H$_2 \longrightarrow$ C$_2$H$^-$ + H$_2$). This may have important observational consequences because the kinetic temperature of the gas is higher at such radii, which would cause the molecule to exist in a higher state of rotational excitation.  Direct comparison of (unresolved) single-dish microwave observations of C$_2$H$^-$ and C$_2$H may be made more difficult by the difference in the distributions and therefore the telescope beam-filling factors of these two species.  Detailed microwave observations of the spatial distributions of C$_2$H$^-$ are clearly required in order to confirm this result and determine whether \pref{eq:c2h2h-} is indeed the dominant production mechanism for this species.

In the new model for the circumstellar envelope of IRC+10216, the respective anion-to-neutral column density ratios for C$_4$H, C$_6$H and C$_8$H are calculated to be 1.4\%, 7.4\% and 4.5\%.  The corresponding observational ratios are 0.024\%, 6.3\% \citep{cer07} and  26\% \citep{rem07}.  As highlighted above for C$_2$H, obtaining accurate observational ratios is hindered by the possibility that the anions and neutrals have different spatial distributions within the CSE.  Nevertheless, it is clear that our model still severely over-estimates the amount of C$_4$H$^-$ compared with C$_4$H \citep[see also][]{her08}.  It may be the case that the C$_4$H radiative electron attachment rate calculated by Herbst is too large.  Alternatively, C$_4$H$^-$ may be destroyed more rapidly than our model currently predicts, for example, by reaction with HCN or C$_2$H$_2$ (cf. Equations (\ref{eq:hcnh-}) and (\ref{eq:c2h2h-})).  In that case, such reactions would have to be much more rapid for C$_4$H$^-$ than for C$_6$H$^-$ and C$_8$H$^-$, which could conceivably result from the different electron binding energies of these species (Herbst, private communication).  Further laboratory and/or theoretical studies are required in order to determine whether the organic anions considered here react with HCN, C$_2$H$_2$ or other possible proton-donating molecules.

\begin{figure}
\centering
\epsfig{file=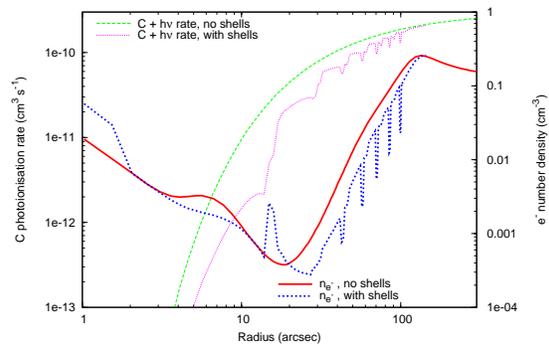, width=0.6\columnwidth, angle=270}
\caption{Electron number density profiles for models with and without density-enhanced shells. The atomic carbon photoionisation rates (C + $h\nu$) for the two models are also plotted.}
\label{fig:e-}
\end{figure}

Anion-to-neutral ratios of the C$_n$ and C$_n$H species are typically slightly greater in the present model compared to the no-shells (NS) model.  This is largely attributable to the increased electron abundances inside the dominant $15''$ shell which leads to increased rates of electron attachment.  \pref{fig:e-} shows the electron abundances as a function of radius. The density-enhanced shells cause increased shielding of the gas from photoionisation which results in electron abundances which are a factor $\sim2$ lower between $r=20''$ and $100''$.  Inside the shells at these radii, contrary to the $15''$ shell, the electron densities are up to an order of magnitude lower than the surrounding CSE.  Beyond $\sim100''$ where the density becomes very low the shells begin to have a negligible effect on the photoionisation rate as shown for atomic carbon in \pref{fig:e-}.  Accordingly, the electron abundances of the models with and without shells converge at this radius.

\subsection{MgNC}
\label{sec:mgnc}

\begin{figure}
\centering
\epsfig{file=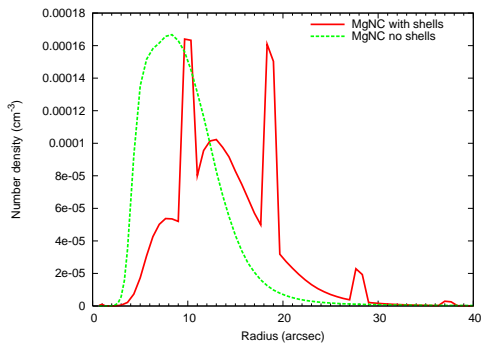, width=0.6\columnwidth, angle=270}
\caption{Modeled MgNC number density profiles with and without density-enhanced shells.}
\label{fig:mgnc}
\end{figure}

A number of metal-containing molecules have been detected in IRC+10216. While some of these, for example NaCl, KCl and AlCl, are expected to be abundant in the LTE region and have spatial distributions that peak on the star, other species, particularly the cyanides and isocyanides such as MgCN, MgNC and AlNC, have shell-like distributions on scales very similar to the cyanopolyynes \citep{ziu02}. Two possibilities exist to account for the presence of these species in the outer envelope: either they form in the gas phase from metals that have not been incorporated into dust in the inner envelope or they have been released from grains by erosion processes in the outer envelope.

In order to test the hypothesis that metal isocyanides are formed by gas-phase chemistry we have investigated the synthesis of MgNC via radiative association reactions between Mg$^+$ and the cyanopolyynes \citep{dun02}:
\be
{\rm Mg^+} + {\rm HC_nN} \longrightarrow {\rm HC_nNMg^+} + {\rm h\nu}
\label{eq:mgnc}
\ee
(with $n = 3,5,7,9$), followed by dissociative recombination reactions of the form:
\be
{\rm HC_nNMg^+} + {\rm e^-} \longrightarrow {\rm MgNC} + {\rm C_{n-1}H}.
\ee

\pref{fig:mgnc} shows the radial abundance (cm$^{-3}$) of MgNC in both the presence and absence of shells. In these calculations we adopt an initial Mg/H$_2$ abundance ratio of 10$^{-5}$ --- the derived abundances and column densities are directly proportional to this ratio. The effect of the density-enhanced shells is, once again, to move the peak of the distribution outward and to concentrate the abundance to the peaks of the gas density, in this case with almost equal peaks in the $15''$ and $29''$ shells. The total MgNC column density is calculated to be 5.7$\times10^{13}$ cm$^{-2}$, compared to observed values of (0.93-5)$\times10^{13}$ cm$^{-2}$ \citep{gue95, hig03}. The dominant formation reaction is with HC$_7$N which, despite its smaller abundance has the largest radiative association rate coefficient (\pref{eq:mgnc}), 6.59$\times10^{-9}(T/300)^{-0.47}$ cm$^3$ s$^{-1}$ \citep{dun02}.  

It was suggested by \citet{dun02} that the inclusion of \pref{eq:mgnc} into chemical models for IRC+10216 might significantly reduce the abundances of the species {\rm HC$_n$N} and help to reconcile the discrepancy between modelled and observed {\rm HC$_n$N} column densities.  In the present model, we find that the reaction of Mg$^+$ with  HC$_5$N, HC$_7$N and HC$_9$N results in only $\sim1$\% reduction in their calculated column densities, which does not significantly improve the match with observations. However, the depletion of these species is dependent (in a roughly linear fashion), on the initial Mg abundance employed.

\section{Discussion}

\begin{deluxetable}{lcccrr}
\tablewidth{0pt}
\tabletypesize{\scriptsize}
\tablecaption{Comparison between column densities from the present model, the present model including no density-enhanced shells (NS), the model by \citet{mil00} (MHB), and observations \label{tab:colds}}
\tablehead{\colhead{Species}&\colhead{Present}&\colhead{NS}&\colhead{MHB}&\colhead{Obs.}&\colhead{Ref.}}
\startdata
C          &2.5e16&  2.7e16&  1.0e16&1.1e16  &    1 \\
C$_2$      &2.6e15&  2.6e15&  9.9e15&7.9e14  &  2   \\
C$_2$H     &5.6e15&  6.3e15&  5.7e15&3-5e15  &  1   \\
CN         &2.2e15&  2.6e15&  1.0e15&1.1e15  &   2  \\
HCO$^+$    &3.1e12&  3.0e12&  2.4e12&3e12    &  1   \\
C$_3$      &3.4e14&  2.6e14&  6.5e14&1e15    &  1   \\
C$_3$H     &1.6e14&  1.2e14&  1.4e14&3-7e13    &  1, 3   \\
C$_3$H$_2$ &1.0e14&  5.6e13&  2.1e13&2e13    &  1   \\
CH$_2$CN   &2.0e13&  1.7e13&  6.9e12&8.4e12  &   4  \\
C$_3$H$_4$ &1.9e12&  5.7e11&  4.0e12&1.6e13  &  4   \\
CH$_3$CN   &3.9e12&  3.6e12&  3.4e12&6-30e12 &  1, 4   \\
C$_4$H     &1.7e15&  1.8e15&  1.0e15&2-9e15  &   1  \\
C$_4$H$^-$ &2.5e13&  2.1e13&    \nodata  &7.1e11  &  5  \\
MgNC       &5.7e13&  6.6e13&    \nodata &0.93-5e13&   6, 7\\
C$_4$H$_2$ &8.0e15&  6.1e15&  2.9e15&3-20e12 &   1  \\
C$_3$N     &7.0e13&  5.7e13&  3.2e14&2-4e14  &  1   \\
C$_3$N$^-$ &9.0e11&  7.7e11&  4.0e11&1.6e12  &  8   \\
HC$_3$N    &6.7e14&  4.8e14&  1.8e15&1-2e15  &  1   \\
CH$_2$CHCN &2.6e11&  6.6e10&  1.1e11&5.1e12  &  4   \\
C$_2$S     &4.1e13&  4.4e13&  3.5e13&9-15e13    &  9, 10   \\
C$_5$      &5.1e14&  6.3e14&  7.5e14&1e14    &   1  \\
C$_5$H     &1.9e14&  2.0e14&  8.7e13&2-50e13 &   1  \\
C$_3$S     &1.5e14&  1.7e14&  6.7e12&6-11e13  &  9, 10   \\
C$_6$H     &1.3e15&  1.5e15&  5.8e14&6.6e13    &  5   \\
C$_6$H$^-$ &9.6e13&  1.1e14&    \nodata  &4.1e12  &  5   \\
C$_5$N     &2.4e14&  2.5e14&  1.4e14&3-6e12    &   1, 3  \\
HC$_5$N    &3.5e15&  2.7e15&  7.1e14&2-3e14  &   1  \\
C$_7$H     &3.7e14&  3.2e14&  4.5e13&1-2e12    &   1, 3  \\
C$_5$S     &1.8e14&  1.7e14&    \nodata  &2.5e13  &  10   \\
C$_8$H     &4.1e14&  4.9e14&  1.1e14&5e12    &   1  \\
C$_8$H$^-$ &2.0e13&  2.1e13&  2.7e13&2e12    &   11  \\
HC$_7$N    &1.4e15&  1.3e15&  2.2e14&1e14    &   1  \\
HC$_9$N    &8.5e14&  8.0e14&  5.8e13&3e13    &   1  \\

\enddata
\tablerefs{(1) see references in Table 5 of \citet{mil00}; (2) \citet{bak97}; (3) \citet{cer00}; (4) \citet{agu08}; (5) \citet{cer07}; (6) \citet{gue95}; (7) \citet{hig03}; (8) \citet{tha08}; (9) \citet{cer87}; (10) \citet{bel93}; (11) \citet{rem07}.}
\end{deluxetable}

Compared to previous chemical models of IRC+10216 \citep[\eg][]{nej87,che93,mil00,bro03,agu08}, the model presented here is unique in the inclusion of density-enhanced shells of gas and dust.  The shells included in our model have physical parameters similar to the dust shells observed by \citet{mau00}. As expected, the modeled column densities differ from those calculated in the model by \citet{mil00} (referred to hereafter as MHB). \pref{tab:colds} gives the column densities for these two models for 33 species for which observational column densities have been published. To highlight the effect of the density-enhanced shells on the chemistry and to permit comparison with models without density-enhanced shells, the column densities from the present model with no density-enhanced shells (referred to as NS), are also given.  The column densities calculated by the three different models are generally in good agreement with observations, especially given the complex morphology of the source.

Notable differences between the MHB and NS models include a substantial reduction in the HC$_3$N/HC$_5$N column density ratio in the new model due to the enhanced photodissociation rate used for HC$_3$N (from the RATE06 database). The C$_3$N abundance is reduced as well because it is produced predominantly from HC$_3$N (\emph{via} an alternative photodissociation channel). The carbon chains C$_6$H, C$_7$H and C$_8$H, the cyanopolyynes HC$_5$N, HC$_7$N and HC$_9$N and also CN are significantly more abundant in the present models than MHB. This is primarily due to the increased initial abundances of the parent species C$_2$H$_2$ and HCN \citep{fon08} to which their chemistry is closely coupled \citep[see][]{mil94}. C$_7$H and C$_8$H are also more abundant as a result of the addition of C$_6$H$^-$ to the chemical network which undergoes associative detachment with H to form C$_6$H$_2$. C$_6$H$_2$ reacts with C to produce C$_7$H or with C$_2$H to produce C$_8$H$_2$, which is photodissociated to C$_8$H. 

Among the 33 chemical species listed in \pref{tab:colds}, 12 show calculated column densities that differ from observations by more than a factor of 10. In the present model, the moderate-to-large-sized hydrocarbons and cyanopolyynes C$_4$H$_2$, C$_5$N, C$_6$H, C$_8$H, and HC$_9$N might be considered to have column densities that are discrepantly greater than the observations.  By reducing the gas-to-dust ratio $G$ to 50 (from the present value of 200), these discrepancies are eliminated due to the retardation of the C$_2$H$_2$ and HCN photochemistry as a consequence of increased dust extinction.  The only species to be detrimentally affected by such a reduction in $G$ are C$_3$N and C$_3$N$^-$, and as mentioned previously, the chemistry involving these species is not completely understood at present. Taking $G=50$ may be too low however, because values from the literature \citep[see Table 5 of][]{men01} are typically in the range 200--1000.  It is plausible that other factors might contribute towards a reduction in the photochemical yields, such as a reduced incident radiation field strength \citep[see][]{mil94} and/or lower initial C$_2$H$_2$ and HCN abundances, to bring the model results into better agreement with observations.  Changes in the rates of the reactions involved in the synthesis of moderate-to-large-sized hydrocarbons and cyanopolyynes could also help to resolve some of the discrepancies between the current model results and observations.  For example, the reactions of the kind C$_2$H + HC$_{2n+1}$N $\longrightarrow$ HC$_{2n+3}$N + H and C$_2$H + C$_n$H$_2$ $\longrightarrow$ C$_{n+2}$H$_2$ + H are crucial for the synthesis of successively larger species in the model, and the rates of these reactions are only known, at best, to within a factor of 2. Similarly, the photodissociation rates of many of the larger species in the model are calculated only from estimated cross-sections.

It has been brought to our attention by the referee that in a recent study of narrow sub-millimetre emission lines from IRC+10216 by \citet{pat08}, a lower limit to the abundance of CS in the inner envelope of 9.3$\times10^{-6}$ was derived. This is significantly greater than our adopted abundance of 4.0$\times10^{-6}$. Thus we have investigated the impact on our model of raising the initial CS abundance to 1.0$\times10^{-5}$. We find that the only species to be appreciably affected are those containing sulphur and those whose chemistries are closely related to the sulphur-containing molecules.  The column densities of HCS, OCS, H$_2$CS and C$_n$S (for $n=1-5$) scale approximately linearly with the initial CS abundance. The C$_3$ and C$_3$H column densities are raised by about 50\%, mainly as a result of the photodissociation of C$_4$S to produce C$_3$ + CS. The OH column density increases by a similar fraction due to the hydrogen-transfer reaction of HCS with O. No other species in \pref{tab:colds} have column densitities that are significantly affected by the increase in the initial CS abundance.

The inclusion of density-enhanced shells in the model has a profound impact on the calculated radial abundance distributions, both through the extinction effect of the shells which modifies the photochemical reaction rates, and through the effects of their increased densities which accelerate the binary chemical reactions. However, the impact of the shells on the total column densities is generally rather small; even though the chemical abundances may vary by over an order of magnitude inside the shells, the fact that the shells are narrow compared to the radial extent of the CSE means that they contribute towards only a modest fraction of the total column densities.  Species whose chemistries are closely linked to the radiation field strength are most affected by the shells, for example C$_3$H$_4$ and CH$_3$CHCN (whose dominant destruction channels are by photodissociation), are shielded from dissociating radiation out to a larger radius by the increased extinction. The enhanced shielding in the envelope also reduces the ionisation rate, resulting in up to an order of magnitude reduction in the electron and C$^+$ column densities at certain radii.

A primary motivation of this study is to examine whether the structure observed by \citet{din08} in HC$_3$N and HC$_5$N maps of IRC+10216 is consistent with their suggestion that the gas and dust are coupled, sharing a similar density distribution.  The inclusion of density-enhanced shells in our model results in peaks in the radial number density distributions of molecules within the shells, which shows that density enhancements in the molecular gas are able to produce small-scale structure in the molecular distributions similar to those observed.

When molecular excitation is taken into account, the emission intensity profiles for C$_2$H, C$_4$H, C$_6$H and HC$_3$N all peak inside the dense shell at $15''$. This finding is consistent with the 3 mm maps of these species made by \citet{gue99} and \citet{luc99} and shows that the presence of dense shells can cause the modeled molecular abundances to peak around the same radius and in a narrow region corresponding to the radius and thickness of the density-enhancement.  The simulated emission maps (\pref{fig:maps}) show a similar picture for C$_2$H and C$_6$H. However, the inner broad intensity maximum of the C$_4$H radial profile results in a $\approx5''$ wide, intense emission ring inside of the density-enhanced $15''$ ring, which is not seen in the \citet{gue99} map.  Possible explanations for this discrepancy might be that the excitation of C$_4$H favours its observation in the narrow shell (contrary to the crude excitation analysis we performed), or that the interferometer used in the observations failed to detect the broader structure of the inner ring. 

In the model, the gas and dust share identically-shaped density profiles which requires that the gas and dust must be dynamically coupled.  This is contrary to the theory \citep[see][]{tro91,mau00} that radiation pressure accelerates the circumstellar dust to a radial drift velocity $\sim2$ \kms\ faster than the gas.  Possible coupling mechanisms may include thermal and turbulent motion of the gas and dust. Future models may need to include a detailed analysis of the coupling between the gas and dust shells.

The shells used in the model represent a gross simplification of the dust shell structure observed in IRC+10216 by \citet{mau00}. However, the results presented here --- that the maximum abundances of species containing carbon chains match the peaks in the circumstellar dust distribution --- should be generally applicable to more complex representations of the density structure of the CSE.  The density-dependent nature of molecular excitation means that detailed excitation calculations are required to determine the relationship between observed emission intensity and molecular abundance distributions.  In our model the temperature distribution of the gas is assumed to be continuous, with the assumption that the density-enhancements are in thermal equilibrium with the surrounding medium.  More detailed observations of the gas and dust will be required to determine if this assumption is realistic.  In addition, accurate collisional and vibrational excitation rates will be required in order to more accurately calculate the rotational excitation of the molecules considered.

\section{Conclusion}

Density-enhanced shells of gas and dust have a significant impact on the calculated radial distributions of molecules in the expanding envelope of IRC+10216.  Based on the suggestion by \citet{din08} that the gas and dust are coupled, we included in a new model for the CSE a set of density-enhancements with parameters based on the dust shell observations by \citet{mau00}. Photochemistry is delayed to greater radii due to increased shielding by the dust shells. Density-enhancements in the gas result in molecular abundances that peak inside the narrow shells, showing that the clumpy structure observed in HC$_3$N and HC$_5$N may be the result of density enhancements in the molecular gas.   The calculated emission intensity profiles for C$_2$H, C$_6$H and HC$_3$N all peak within a narrow band about 15$''$ from the central star, which is consistent with detailed emission maps of these species.  The emission profile for C$_4$H has a broad maximum around $8''$ that is not present in observed maps, which may be indicative of a need for further improvements in the model.

The ethynyl anion C$_2$H$^-$ and the nitrile anions C$_{2n-1}$N$^-$ (for $n=1, 2, 3, 4$), have been calculated to reach observable abundances in the CSE. For the smaller nitrile anions for which radiative electron attachment is slow (\ie\ CN$^-$ and C$_3$N$^-$), this result is uncertain due to the unknown product branching ratios in the associative electron detachment channels of the reactions N + C$_m^-$ and N + C$_m$H$^-$. Further laboratory measurements (including up to at least $m=10$), will be required in order to confirm the importance of these reactions in anion astrochemistry. C$_2$H$^-$ is predicted to be produced in abundance in the inner CSE (at much smaller radii than the other anions in the model), as a product of the reaction of H$^-$ with C$_2$H$_2$.  Other possible proton transfer reactions of this kind (\ie\ XH + Y$^-$ $\longrightarrow$ X$^-$ + YH, for the anions listed in \pref{tab:anioncolds}), also need to be considered for possible inclusion in future anion chemical networks.

Our chemical models produce MgNC with a peak abundance in the outer envelope. Dependent on the initial Mg abundance used, the observed MgNC column density matches observation, which shows that gas-phase chemistry is a viable route to the formation of this species.

\acknowledgments
We gratefully acknowledge Veronica Bierbaum and Eric Herbst for their contributions to this work regarding the calculation of branching ratios for reactions between anions and nitrogen atoms. MAC thanks QUB for financial support. Astrophysics at QUB is supported by a grant from the STFC.

\end{document}